\documentclass[preprint,showpacs,preprintnumbers,amsmath,amssymb]{revtex4}

\usepackage{graphicx}
\usepackage{dcolumn}
\usepackage{bm}

\begin{document}

\preprint{APS/123-QED}

\title{Quasiquartet CEF ground state with possible quadrupolar ordering in the tetragonal compound YbRu$_{2}$Ge$_{2}$}
\author{H. S. Jeevan}
\email{jeevan@cpfs.mpg.de}
\author{C. Geibel}
\affiliation{Max Planck Institute for Chemical Physic of Solids,
N\"othnitzer Str. 40, D 01187 Dresden, Germany}
\author{Z. Hossain}
\affiliation{ Department of Physics, Indian Institute of Technology,
Kanpur 208016, India}
\date{\today}
\begin{abstract}
We have investigated the magnetic properties of YbRu$_{2}$Ge$_{2}$
by means of magnetic susceptibility $\chi$(T), specific heat C(T)
and electrical resistivity $\rho$(T) measurements performed on flux
grown single crystals. The Curie-Weiss behavior of $\chi$(T) along
the easy plane, the large magnetic entropy at low temperatures and
the weak Kondo like increase in $\rho$(T) proves a stable trivalent
Yb state. Anomalies in C(T), $\rho$(T) and $\chi$(T) at T$_{0}$ =
10.2 K, T$_{1}$ = 6.5 K and T$_{2}$ = 5.7 K evidence complex
ordering phenomena, T$_{0}$ being larger than the highest Yb
magnetic ordering temperature found up to now. The magnetic entropy
just above T$_{0}$ amounts to almost Rln4, indicating that the
crystal electric field (CEF) ground state is a quasiquartet instead
of the expected doublet. The behavior at T$_{0}$ is rather unusual
and suggest that this transition is related to quadrupolar ordering,
being a consequence of the CEF quasiquartet ground state. The
combination of a quasiquartet CEF ground state, a high ordering
temperature, and the relevance of quadrupolar interactions makes
YbRu$_{2}$Ge$_{2}$ a rather unique system among Yb based compounds.
\end{abstract}

\maketitle

In intermetallic compounds based on Ce or Yb, the instability of the
\emph{f}-shell of these elements allows them to be tuned from a
magnetic to a non-magnetic state by changing the chemical
composition or by applying pressure. At the crossover from the
non-magnetic to the magnetic state, one observes unusual properties
like the formation of heavy fermions, the onset of unconventional
superconductivity or strong deviation from the Fermi-Liquid behavior
usually expected in a metal. A large part of the research in this
field was performed on CeT$_{2}$X$_{2}$ compounds (T = transition
metals, X = Si and Ge) crystallizing in the ThCr$_{2}$Si$_{2}$ or a
related structure type, because some of these compounds are very
close to this crossover. Two prominent examples are
CeRu$_{2}$Si$_{2}$ \cite{1,2} and CeRh$_{2}$Si$_{2}$ \cite{3}, the
former is just on the non-magnetic side and shows an unconventional
metamagnetic transition from a delocalized to a localized
\emph{f}-state, while the later one, although being just on the
magnetic side of the crossover, has the highest antiferromagnetic
ordering temperature among all Ce-compounds. While all the
CeT$_{2}$X$_{2}$ compounds have now been thoroughly investigated,
much less studies were performed on the Yb-based homologues. For the
Yb-compounds with T = Ru, Os, Rh, Ir, only little or nothing is know
about their physical properties. YbRh$_{2}$Si$_{2}$  \cite{4,5} was
investigated only quite recently, and was found to be located
extremely close to the quantum critical point connected with the
onset of magnetic ordered state, which leads to very interesting
properties and makes this compound one of the most interesting ones
in the field of quantum phase transitions. The first investigation
of YbIr$_{2}$Si$_{2}$ \cite{6} has just been published, it is a
heavy fermion system just on the non-magnetic side of the critical
point. In search for further interesting Yb-based compounds we
synthesized YbRu$_{2}$Ge$_{2}$ and investigated its physical
properties. To the best of our knowledge only structural data have
been reported up to now  \cite{7}. Our results revealed a stable
trivalent Yb state, rather complex ordering phenomena with 3
successive transitions at T$_{0}$ = 10.2 K, T$_{1}$ = 6.5 K and
T$_{2}$ = 5.7 K, and, to our surprise, a quasiquartet crystal field
ground state which is a unique situation among YbT$_{2}$X$_{2}$
compound. The behavior observed at T$_{0}$ suggest this transition
to be a quadrupolar one, being a consequence of the quasiquartet CEF
ground state. The combination of a quasiquartet CEF ground state, a
high Yb ordering temperature and the likely relevance of quadrupolar
interactions makes YbRu$_{2}$Ge$_{2}$ a unique system among Yb-based
compounds.

Polycrystalline samples were prepared by a sintering method, while
single crystals were grown from a Sn or In flux  \cite{8}. XR-powder
diffraction pattern of the polycrystals confirmed the
ThCr$_{2}$Si$_{2}$ (I4/mmm) structure type, a few weak peaks could
not be indexed indicating foreign phases with an amount $<$ 5$\%$.
 Electron probe microanalysis of the single crystals and X-ray powder
diffraction pattern taken from crushed single crystals showed that
some of the single crystals were single phase YbRu$_{2}$Ge$_{2}$,
while others had little impurity phase $<$ 5$\%$ at the surfaces.
The lattice parameters of our single crystals, \emph{a} = 4.2116
(10)\AA  and \emph{c} = 9.7545(20)\AA  were slightly different from
those obtained in our polycrystals, \emph{a} = 4.2105 (10)\AA and
\emph{c} = 9.7567 (20)\AA, and differed significantly from those
reported in the literature, \emph{a} = 4.203 (4)\AA  and \emph{c} =
9.763 (9)\AA  \cite{7}. This suggests the existence either of a
significant homogeneity range (likely along the Ge-Ru line) or of
Ge-Ru disorder, changes in composition or ordering leading to a
decrease of the lattice parameter \emph{a} and a much weaker
increase of \emph{c}. The magnetic susceptibility in the temperature
range 2K-300K and in an applied field from 1 T to 5 T was measured
in a commercial Quantum Design SQUID magnetometer (MPMS). The
specific-heat and the resistivity were determined in a commercial
PPMS (Quantum Design) equipment using the relaxation method and a
four probe ac-technique, respectively. All physical properties
reported here were measured on the best single crystals which had no
impurities as revealed by optical microscope and electron probe
microanalysis as well as a high residual resistivity ratio. The
polycrystalline samples showed similar physical properties, but with
broader transitions.
\begin{figure}
\includegraphics[width=8.5cm]{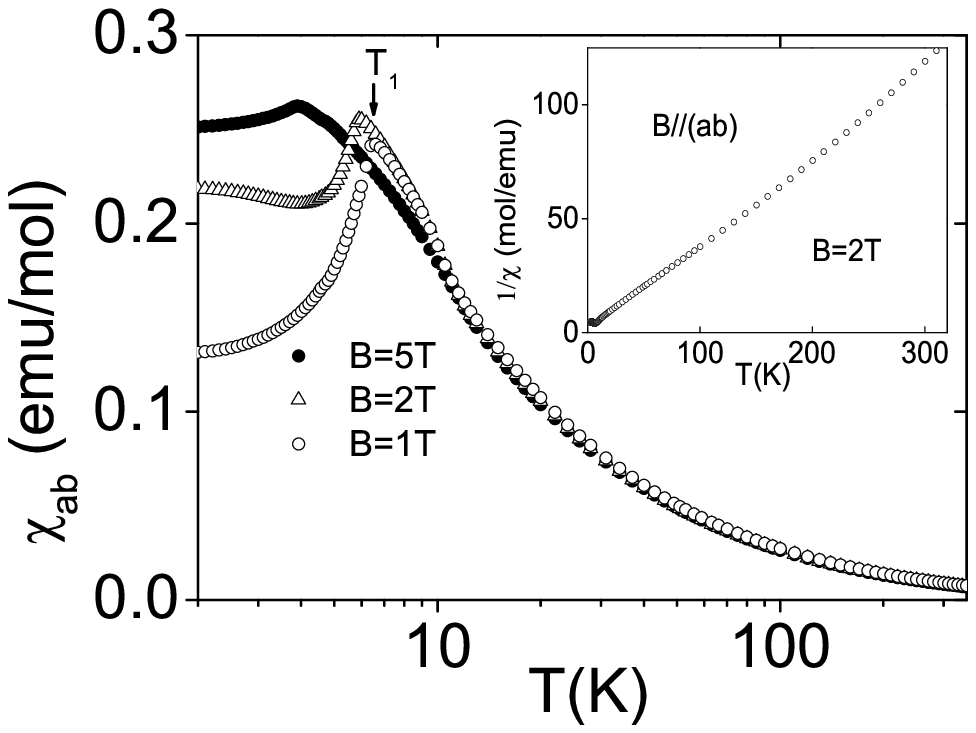}
\caption{\label{fig 1} Temperature dependence of magnetic
susceptibility of YbRu$_{2}$Ge$_{2}$ for magnetic fields B= 1T, 2T
and 5T applied along basal plane. Inset: T-dependence of
1/$\chi$$_{ab}$ for B= 2T.}
\end{figure}
\begin{figure}
\includegraphics[width=8.5cm]{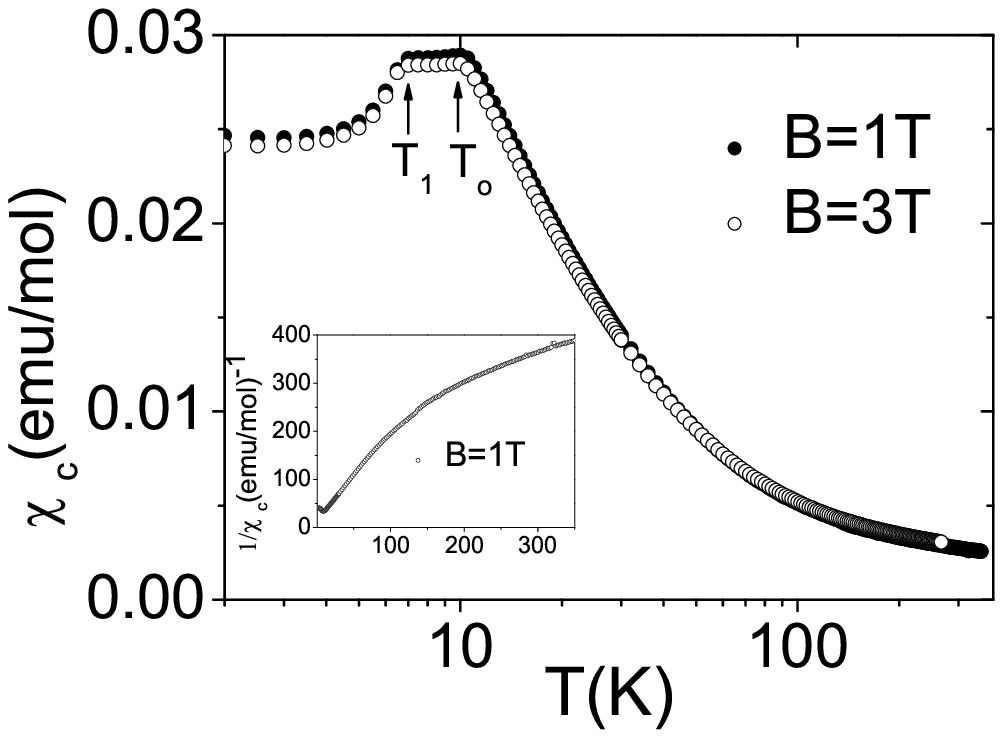}
\caption{\label{fig 2} Temperature dependence of magnetic
susceptibility of YbRu$_{2}$Ge$_{2}$ for magnetic fields B= 1T and
3T applied along \emph{c} axis. Inset  T-dependence of
1/$\chi$$_{c}$ for B= 1T. }
\end{figure}

The magnetic susceptibility gives a first evidence for a trivalent
Yb state. $\chi$(T) is strongly anisotropic, being much larger for
field along the basal plane (Fig.1) than field along the \emph{c}
axis (Fig .2). For field along the basal plane, $\chi$$_{ab}$
follows rather nicely a Curie-Weiss law from 20 K up to room
temperature. The slight curvature in the 1/$\chi$$_{ab}$(T) versus T
plot (inset of Fig.1) at low and high temperatures can be attributed
to crystal field effects and a very small T independent diamagnetic
contribution, respectively. The value of the effective moment
between 20 and 300 K, 4.5 $\mu$$_{B}$, is very close to the value
expected for a trivalent Yb state, 4.54 $\mu$$_{B}$. A pronounced
drop of $\chi$$_{ab}$(T) below T$_{1}$ = 6.5 K evidence a transition
to a magnetically ordered state, while no anomaly is visible around
10 K, the temperature of a pronounced anomaly in C(T) (see below).
The susceptibility for field along the c axis, $\chi$$_{c}$, is one
order of magnitude lower than for field along the basal plane (Fig.
2). The 1/$\chi$$_{c}$(T) versus T curve shows a pronounced negative
curvature. As a result the slope at 300 K is still slightly smaller
( 20$\%$) than that expected for a free Yb$^3$$^+$state. Such a
pronounced curvature in 1/$\chi$(T) for field along the hard axis
can be attributed to a rather large overall crystal field splitting.
In the related compounds YbIr$_{2}$Si$_{2}$ \cite{9} and
YbRh$_{2}$Si$_{2}$  \cite{10}, the highest excited CEF level is
indeed above 400 K. At low temperatures, $\chi$$_{c}$(T) shows a
significant change of slope at T$_{0}$ = 10.2 K, followed by a very
pronounced decrease below T$_{1}$.

\begin{figure}
\includegraphics[width=8.5cm]{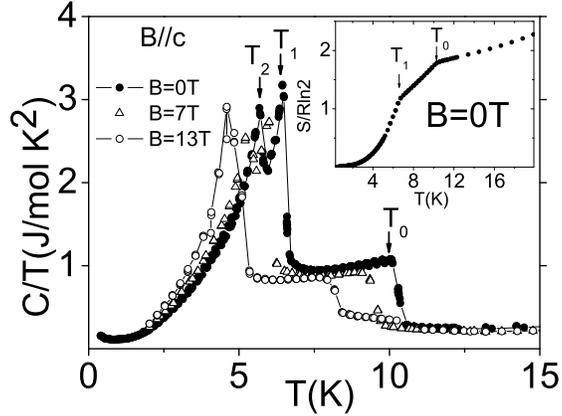}
\caption{\label{fig 3}  Temperature dependence of the heat capacity
of YbRu$_{2}$Ge$_{2}$, in a plot C/T versus T, for different fields
applied along the c axis. The inset shows the temperature dependence
of the entropy.}
\end{figure}

The results of the specific heat measurements (Fig.3) confirm the
trivalent Yb state and the presence of several phase transitions at
low temperatures. The plot C/T versus T (Fig.3) evidences three
distinct transitions: a large mean field like anomaly at T$_{0}$ =
10.2 K and two well resolved and sharp $\lambda$ type anomalies at
T$_{1}$ = 6.5 K and T$_{2}$ = 5.7 K. All these three transitions
were well reproduced in different single crystals, showing that they
are intrinsic. The large size of the anomalies confirms a trivalent
Yb state with a localized 4\emph{f} hole which undergoes ordering at
low temperatures. Below T$_{2}$, the decreases of C/T is
proportional to T$^2$, as expected for an antiferromagnet, but below
2 K the temperature dependence evolves towards a larger exponent,
indicating an exponential suppression of the magnetic excitations
related to the presence of an anisotropy gap. Finally, below 1 K,
C/T merges into a constant Sommerfeld coefficient $\gamma$ $\approx$
100 mJ/K$^2$mol. The slight upturn below 0.8 K can be attributed to
a nuclear Schottky contribution.
\begin{figure}
\includegraphics[width=8.5cm]{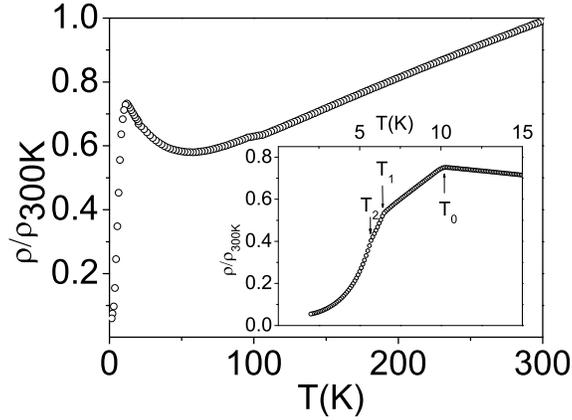}
\caption{\label{fig 4}  Temperature dependence of the resistivity
normalized to its value at 300 K. The inset shows $\rho$(T) from 2 -
15 K. Kinks marked by arrows correspond to transitions at T$_{0}$,
T$_{1}$ and T$_{2}$.}
\end{figure}

The temperature dependence of the normalized resistivity is shown in
Fig.4. The resistivity was measured for current in the basal plane.
The resistivity ratio $\rho$(300K)/ $\rho$(2K) = 22 is an indicator
for the good quality of the sample. The resistivity linearly
decreases with temperature from room temperature down to 70 K, shows
a minimum around 50 K below which the resistivity increases with
decreasing temperature. This increase was sample dependent, in
contrast to all other features in $\rho$(T) which were very
reproducible. We tentatively attribute this increas to some weak
Kondo type interaction. At 10.2 K $\rho$(T) exhibits a sharp break
in the slope, from a negative one at T $>$ T$_{0}$ to a positive one
at T $<$ T$_{0}$. The slope in $\rho$(T) increase further very
strongly at T$_{1}$ = 6.5 K and only slightly more at T$_{2}$= 5.7
K. Thus all the three transitions are also visible in the
resistivity. Interestingly, it turns out that there is a quite good
correspondence between C(T), d$\rho$(T)/dT and d$\chi$$_{c}$(T)/dT
in the temperature range of the transitions, between 12 K and 2 K.
The decrease in $\rho$(T) below the transition, especially below
T$_{1}$, can be attributed to the freezing out of spin disorder
scattering.
\begin{figure}
\includegraphics[width=8.5cm]{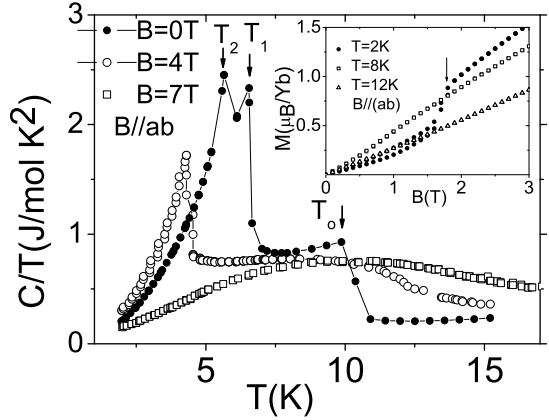}
\caption{\label{fig 5}  This figure shows the strong dependence of
the specific heat on field applied along the Basal plane in a plot
C/T versus T for B= 0T, 4T, 7T. Inset shows the metamagnetic
transition (arrow) at B=1.8T observed in a magnetization M versus B
plot at T= 2K for field along the basal plane.}
\end{figure}

The effects of a magnetic field strongly depend on its direction. A
field along the hard direction has only little influence: The size
and the shape of the anomalies in the specific heat (Fig.3), in the
susceptibility (Fig.2) and in the resistivity (not shown) remains
unaffected, the only change being the merging of T$_{1}$  and
T$_{2}$  for field larger than 10 T. All the transition temperatures
decreases only slightly with increasing field, by just 20$\%$ in 13
T, the shifts being roughly proportional to B$^2$. The effect of a
field along the easy plane is much stronger and differs strongly for
T$_{0}$ and T$_{1}$. T$_{1}$ is shifted to lower temperature with
increasing field (inset of Fig.1), down to 4 K at B = 5 T and to
below 2 K at 7T (Fig.5), the shift being also roughly proportional
to B$^2$. Further on, the decrease observed in $\chi$$_{ab}$(T)
below T$_{1}$ is strongly suppressed in field larger than 1 T. A
magnetization curve (Inset of Fig.5) reveals that this change in
$\chi$(T) is connected with a small metamagnetic transition, visible
as a jump $\Delta$M of only 0.5$\mu$$_{B}$/Yb in the M versus B plot
for T $<$ T$_{1}$. In contrast M(B) display only a very weak
curvature for T$_{1}$ $<$ T $<$ T$_{0}$ and is linear for T $>$
T$_{0}$. This suggests a spin flop like transition to occur at B =
1.8 T for T $<$ T$_{1}$. In contrast, the upper transition T$_{0}$
shifts to higher temperatures with increasing field, up to 12 K at B
= 4 T, and broadens (Fig.5). At B = 7 T the specific heat reveal
only a broad, Schottky like anomaly, no clear transition. The origin
of this unusual behavior will be discussed below.

Up to now, except for the unusual behavior at T$_{0}$,
YbRu$_{2}$Ge$_{2}$ seems to present the classical behavior of a
trivalent Yb-compound. The surprise came when we looked at the
magnetic entropy S(T), which we calculated by integrating the
measured C(T)/T (inset of Fig.3). Because LuRu$_{2}$Ge$_{2}$ does
not form, it was not possible to determine and subtract the phonon
contribution to the C(T). However, for the calculation of S(T) in
the temperature range considered here (T $<$ 12 K), the phonon
contribution can be safely neglected because its contribution is
negligible. As an example, the total entropy of the non-magnetic
compound LuRh$_{2}$Si$_{2}$ at 12 K amounts to 0.17 J/molK, less
than 2$\%$ of the total entropy we determined for YbRu$_{2}$Ge$_{2}$
at the same temperature. In a tetragonal environment, the crystal
field splits the J = 7/2 state of Yb$^3$$^+$ into four Kramer
doublets, with an energy spacing usually larger than 50 K. Thus only
the lowest doublet is relevant for the magnetic properties at low
temperatures and one expects S(T) close to Rln2 slightly above
T$_{N}$. Our surprising result is that the magnetic entropy of
YbRu$_{2}$Ge$_{2}$ just above the highest transition T$_{0}$ is much
larger, it almost reaches Rln4. This result, which was reproduced
with different samples, proves that in YbRu$_{2}$Ge$_{2}$, by
accident, the first excited crystal field doublet is almost
degenerated with the ground state doublet, the excitation energy
being less than 10 K. After our observation we looked in the
literature  \cite{11}, and found that for the homologue and
isoelectronic compound YbRu$_{2}$Si$_{2}$, a CEF calculation based
on an extrapolation of the CEF scheme of other RRu$_{2}$Si$_{2}$
compounds (R = Rare earth) postulated a small excitation energy (25
K) for the first excited CEF level. However, there is up to now no
experimental confirmation of such a low level splitting in
YbRu$_{2}$Si$_{2}$. Thus our investigation reveals a very unusual
quasiquartet CEF ground state in YbRu$_{2}$Ge$_{2}$, which is unique
among YbT$_{2}$X$_{2}$ compounds.

With the presence of a quasiquartet CEF ground state, quadrupolar
ordering became relevant and has to be considered. For the
transition T$_{1}$, the situation is rather obvious: the strong
decrease of $\chi$$_{ab}$(T) and $\chi$$_{c}$(T) below T$_{1}$
indicate that this transition corresponds to antiferromagnetic
ordering. As a consequence, the transition at T$_{2}$ is likely
related to a change of the magnetic structure. In contrast, for
T$_{0}$ the situation is much less clear. At first we note that
T$_{0}$ is larger than the highest magnetic ordering temperature
reported up to now in intermetallic Yb compounds, T$_{N}$ = 7.5 K in
Yb$_{3}$Cu$_{4}$Ge$_{4}$ \cite{12}. The absence of a visible anomaly
in the easy plane susceptibility at T$_{0}$, despite a large mean
field like anomaly in C(T) and a weak anomaly in the susceptibility
along the hard direction, is unusual for a pure magnetic ordering.
Also the increase of T$_{0}$ for field along the easy direction is
not expected for an antiferromagnetic transition in a
three-dimensional system. In contrast, these results correspond to
the behavior expected for quadrupolar ordering. As an example, our
observations in YbRu$_{2}$Ge$_{2}$ are almost identical to those
reported for TmAu$_{2}$  \cite{13}, where the upper transition was
revealed to be ferroquadrupolar ordering. The behavior we observe at
T$_{0}$ and the similarities with TmAu$_{2}$ strongly suggest that
the transition at T$_{0}$ in YbRu$_{2}$Ge$_{2}$ corresponds to
quadrupolar ordering \cite{14}. While for some of the rare earth
quadrupolar ordering is quite common, it is rather exceptional among
intermetallic Yb compounds. The only well established example is
YbSb  \cite{15}, which shows a mixed type AFQ at 5.0 K. A
quadrupolar ordering has recently be suggested to occur in
YbAl$_{3}$C$_{3}$ \cite{16}, but this claim remains speculative
because of an extremely large quadrupolar transition temperature
T$_{Q}$ = 80 K, exceeding all previously reported quadrupolar
ordering temperatures by a factor of two, and the hexagonal symmetry
of the structure, which should lead to a doublet CEF ground state
and thus be rather unfavorable for quadrupolar ordering. Thus
YbRu$_{2}$Ge$_{2}$ likely presents a unique type of ordering among
Yb-compounds. Our results might also have some consequences for the
interpretation of the unusual properties of YbRh$_{2}$Si$_{2}$.

In summary, we have grown single crystals of YbRu$_{2}$Ge$_{2}$ and
investigated the physical properties of this compound by means of
susceptibility, specific heat and resistivity measurements. The
susceptibility is strongly anisotropic, being much larger for field
in the basal plane than along the \emph{c} direction. For field
along the easy plane $\chi$$_{ab}$(T) follows a Curie-Weiss law with
an effective moment close to the value for free Yb$^3$$^+$, while
for field along the hard direction the curve 1/$\chi$$_{c}$(T)
versus T shows a strong negative curvature indicating a large
overall CEF splitting. The temperature dependence of the resistivity
follows a standard metallic behavior above 50 K and shows a weak
Kondo type increase below 50 K. At lower temperatures, anomalies in
C(T), $\rho$(T) and $\chi$(T) evidence three successive phase
transitions at T$_{0}$ = 10.2 K, T$_{1}$ = 6.5 K and T$_{2}$ = 5.7
K, T$_{0}$ being larger than the highest Yb-magnetic ordering
temperature observed up to now in intermetallic Yb compounds. Just
above T$_{0}$, the magnetic entropy calculated from the specific
heat reaches almost Rln4 instead of Rln2 expected from the CEF
ground state doublet. The large anisotropy of the susceptibility,
the Curie Weiss behavior of $\chi$(T) along the easy plane, the
large magnetic entropy collected at low temperatures and the
weakness of the Kondo like increase in $\rho$(T) demonstrate that Yb
is in a stable trivalent state. S(T $\geq$ T$_{0}$) $\approx$ Rln4
proves that the energy of the first CEF excited doublet is lower
than 10 K, leading to a quasiquartet CEF ground state, a unique
situation among YbT$_{2}$X$_{2}$ compounds. The shape of the
anomalies at T$_{0}$ in $\chi$(T) and its behavior in a magnetic
field are unusual for magnetic ordering, but very similar to those
reported at the ferroquadrupolar ordering in TmAu$_{2}$. In view of
the quasiquartet CEF ground state, this strongly suggests the
transition at T$_{0}$ to be quadrupolar ordering. In contrast the
strong decrease of $\chi$(T) at T$_{1}$ indicates antiferromagnetic
ordering, while the transition T$_{2}$ is likely related to a change
in the magnetic structure. The combination of a quasiquartet CEF
ground state, a high ordering temperature and the likely relevance
of quadrupolar interactions makes YbRu$_{2}$Ge$_{2}$ a very
interesting system among Yb-based compounds. Neutron diffraction and
muSR experiments are now in progress in order to reveal the nature
of the different transitions.

 The authors would like to thank U. Burkhardt and P. Scheppen for
chemical analysis of the samples

\begin{thebibliography}{16}
\expandafter\ifx\csname
natexlab\endcsname\relax\def\natexlab#1{#1}\fi
\expandafter\ifx\csname bibnamefont\endcsname\relax
  \def\bibnamefont#1{#1}\fi
\expandafter\ifx\csname bibfnamefont\endcsname\relax
  \def\bibfnamefont#1{#1}\fi
\expandafter\ifx\csname citenamefont\endcsname\relax
  \def\citenamefont#1{#1}\fi
\expandafter\ifx\csname url\endcsname\relax
  \def\url#1{\texttt{#1}}\fi
\expandafter\ifx\csname urlprefix\endcsname\relax\def\urlprefix{URL
}\fi \providecommand{\bibinfo}[2]{#2}
\providecommand{\eprint}[2][]{\url{#2}}

\bibitem[{\citenamefont{Haen et~al.}(1987)\citenamefont{Haen, Flouquet,
  Lapierre, Lejay, and Remenyi}}]{1}
\bibinfo{author}{\bibfnamefont{P.}~\bibnamefont{Haen}},
  \bibinfo{author}{\bibfnamefont{J.}~\bibnamefont{Flouquet}},
  \bibinfo{author}{\bibfnamefont{F.}~\bibnamefont{Lapierre}},
  \bibinfo{author}{\bibfnamefont{P.}~\bibnamefont{Lejay}}, \bibnamefont{and}
  \bibinfo{author}{\bibfnamefont{G.}~\bibnamefont{Remenyi}},
  \bibinfo{journal}{J. Low Temp. Phys} \textbf{\bibinfo{volume}{67}},
  \bibinfo{pages}{779} (\bibinfo{year}{1987}).

\bibitem[{\citenamefont{Sakakibara et~al.}(1995)\citenamefont{Sakakibara,
  Tayama, Matsuhira, Mitamura, , Amitsuka, Maezawa, and Onuki}}]{2}
\bibinfo{author}{\bibfnamefont{T.}~\bibnamefont{Sakakibara}},
  \bibinfo{author}{\bibfnamefont{T.}~\bibnamefont{Tayama}},
  \bibinfo{author}{\bibfnamefont{K.}~\bibnamefont{Matsuhira}},
  \bibinfo{author}{\bibfnamefont{H.}~\bibnamefont{Mitamura}}, ,
  \bibinfo{author}{\bibfnamefont{H.}~\bibnamefont{Amitsuka}},
  \bibinfo{author}{\bibfnamefont{K.}~\bibnamefont{Maezawa}}, \bibnamefont{and}
  \bibinfo{author}{\bibfnamefont{Y.}~\bibnamefont{Onuki}},
  \bibinfo{journal}{Phys. Rev. B} \textbf{\bibinfo{volume}{51}},
  \bibinfo{pages}{R12030} (\bibinfo{year}{1995}).

\bibitem[{\citenamefont{Movshovich et~al.}(1996)\citenamefont{Movshovich, Graf,
  Mandrus, Thompson, Smith, and Fisk}}]{3}
\bibinfo{author}{\bibfnamefont{R.}~\bibnamefont{Movshovich}},
  \bibinfo{author}{\bibfnamefont{T.}~\bibnamefont{Graf}},
  \bibinfo{author}{\bibfnamefont{D.}~\bibnamefont{Mandrus}},
  \bibinfo{author}{\bibfnamefont{J.~D.} \bibnamefont{Thompson}},
  \bibinfo{author}{\bibfnamefont{J.~L.} \bibnamefont{Smith}}, \bibnamefont{and}
  \bibinfo{author}{\bibfnamefont{Z.}~\bibnamefont{Fisk}},
  \bibinfo{journal}{Phys. Rev. B} \textbf{\bibinfo{volume}{53}},
  \bibinfo{pages}{8241} (\bibinfo{year}{1996}).

\bibitem[{\citenamefont{Trovarelli et~al.}(2000)\citenamefont{Trovarelli,
  Geibel, Mederle, Langhammer, Grosche, Gegenwart, Lang, Sparn, and
  Steglich}}]{4}
\bibinfo{author}{\bibfnamefont{O.}~\bibnamefont{Trovarelli}},
  \bibinfo{author}{\bibfnamefont{C.}~\bibnamefont{Geibel}},
  \bibinfo{author}{\bibfnamefont{S.}~\bibnamefont{Mederle}},
  \bibinfo{author}{\bibfnamefont{C.}~\bibnamefont{Langhammer}},
  \bibinfo{author}{\bibfnamefont{F.~M.} \bibnamefont{Grosche}},
  \bibinfo{author}{\bibfnamefont{P.}~\bibnamefont{Gegenwart}},
  \bibinfo{author}{\bibfnamefont{M.}~\bibnamefont{Lang}},
  \bibinfo{author}{\bibfnamefont{G.}~\bibnamefont{Sparn}}, \bibnamefont{and}
  \bibinfo{author}{\bibfnamefont{F.}~\bibnamefont{Steglich}},
  \bibinfo{journal}{Phys. Rev. Lett} \textbf{\bibinfo{volume}{85}},
  \bibinfo{pages}{626} (\bibinfo{year}{2000}).

\bibitem[{\citenamefont{Paschen et~al.}(2004)\citenamefont{Paschen,
  L{\"{u}}hmann, Wirth, Gegenwart, Trovarelli, Geibel, Steglich, Coleman, and
  Si}}]{5}
\bibinfo{author}{\bibfnamefont{S.}~\bibnamefont{Paschen}},
  \bibinfo{author}{\bibfnamefont{T.}~\bibnamefont{L{\"{u}}hmann}},
  \bibinfo{author}{\bibfnamefont{S.}~\bibnamefont{Wirth}},
  \bibinfo{author}{\bibfnamefont{P.}~\bibnamefont{Gegenwart}},
  \bibinfo{author}{\bibfnamefont{O.}~\bibnamefont{Trovarelli}},
  \bibinfo{author}{\bibfnamefont{C.}~\bibnamefont{Geibel}},
  \bibinfo{author}{\bibfnamefont{F.}~\bibnamefont{Steglich}},
  \bibinfo{author}{\bibfnamefont{P.}~\bibnamefont{Coleman}}, \bibnamefont{and}
  \bibinfo{author}{\bibfnamefont{Q.}~\bibnamefont{Si}},
  \bibinfo{journal}{Nature} \textbf{\bibinfo{volume}{432}},
  \bibinfo{pages}{881} (\bibinfo{year}{2004}).

\bibitem[{\citenamefont{Hossain et~al.}(2005)\citenamefont{Hossain, Geibel,
  Weickert, Radu, Tokiwa, Jeevan, Gegenwart, and Steglich}}]{6}
\bibinfo{author}{\bibfnamefont{Z.}~\bibnamefont{Hossain}},
  \bibinfo{author}{\bibfnamefont{C.}~\bibnamefont{Geibel}},
  \bibinfo{author}{\bibfnamefont{F.}~\bibnamefont{Weickert}},
  \bibinfo{author}{\bibfnamefont{T.}~\bibnamefont{Radu}},
  \bibinfo{author}{\bibfnamefont{Y.}~\bibnamefont{Tokiwa}},
  \bibinfo{author}{\bibfnamefont{H.}~\bibnamefont{Jeevan}},
  \bibinfo{author}{\bibfnamefont{P.}~\bibnamefont{Gegenwart}},
  \bibnamefont{and} \bibinfo{author}{\bibfnamefont{F.}~\bibnamefont{Steglich}},
  \bibinfo{journal}{Phys. Rev. B} \textbf{\bibinfo{volume}{72}},
  \bibinfo{pages}{94411} (\bibinfo{year}{2005}).

\bibitem[{\citenamefont{Francois et~al.}(2005)\citenamefont{Francois,
  Venturini, Mareche, Malaman, and Roques}}]{7}
\bibinfo{author}{\bibfnamefont{M.}~\bibnamefont{Francois}},
  \bibinfo{author}{\bibfnamefont{G.}~\bibnamefont{Venturini}},
  \bibinfo{author}{\bibfnamefont{J.~F.} \bibnamefont{Mareche}},
  \bibinfo{author}{\bibfnamefont{B.}~\bibnamefont{Malaman}}, \bibnamefont{and}
  \bibinfo{author}{\bibfnamefont{B.}~\bibnamefont{Roques}},
  \bibinfo{journal}{J.Less-comm Metal} \textbf{\bibinfo{volume}{113}},
  \bibinfo{pages}{231} (\bibinfo{year}{2005}).

\bibitem[{\citenamefont{Jeevan et~al.}(To be published)\citenamefont{Jeevan,
  Hossain, and Geibel}}]{8}
\bibinfo{author}{\bibfnamefont{H.~S.} \bibnamefont{Jeevan}},
  \bibinfo{author}{\bibfnamefont{Z.}~\bibnamefont{Hossain}}, \bibnamefont{and}
  \bibinfo{author}{\bibfnamefont{C.}~\bibnamefont{Geibel}} (\bibinfo{year}{To
  be published}).

\bibitem[{\citenamefont{Hiess et~al.}(2005)\citenamefont{Hiess, Stockert, Koza,
  Hossain, and Geibel}}]{9}
\bibinfo{author}{\bibfnamefont{A.}~\bibnamefont{Hiess}},
  \bibinfo{author}{\bibfnamefont{O.}~\bibnamefont{Stockert}},
  \bibinfo{author}{\bibfnamefont{M.~M.} \bibnamefont{Koza}},
  \bibinfo{author}{\bibfnamefont{Z.}~\bibnamefont{Hossain}}, \bibnamefont{and}
  \bibinfo{author}{\bibfnamefont{C.}~\bibnamefont{Geibel}},
  \bibinfo{journal}{Physica B,} \textbf{\bibinfo{volume}{in press}}
  (\bibinfo{year}{2005}).

\bibitem[{\citenamefont{Stockert et~al.}(2005)\citenamefont{Stockert, Koza,
  Ferstl, Murani, Geibel, and Steglich}}]{10}
\bibinfo{author}{\bibfnamefont{O.}~\bibnamefont{Stockert}},
  \bibinfo{author}{\bibfnamefont{M.~M.} \bibnamefont{Koza}},
  \bibinfo{author}{\bibfnamefont{J.}~\bibnamefont{Ferstl}},
  \bibinfo{author}{\bibfnamefont{A.~P.} \bibnamefont{Murani}},
  \bibinfo{author}{\bibfnamefont{C.}~\bibnamefont{Geibel}}, \bibnamefont{and}
  \bibinfo{author}{\bibfnamefont{F.}~\bibnamefont{Steglich}},
  \bibinfo{journal}{Physica B,} \textbf{\bibinfo{volume}{in press}}
  (\bibinfo{year}{2005}).

\bibitem[{\citenamefont{Michalski et~al.}(2003)\citenamefont{Michalski, Blaut,
  and Radwanski}}]{11}
\bibinfo{author}{\bibfnamefont{R.}~\bibnamefont{Michalski}},
  \bibinfo{author}{\bibfnamefont{A.}~\bibnamefont{Blaut}}, \bibnamefont{and}
  \bibinfo{author}{\bibfnamefont{R.}~\bibnamefont{Radwanski}},
  \bibinfo{journal}{Acta Physica Polonica B} \textbf{\bibinfo{volume}{34}},
  \bibinfo{pages}{1565} (\bibinfo{year}{2003}).

\bibitem[{\citenamefont{Dhar et~al.}(2002)\citenamefont{Dhar, Singh, Bonville,
  Mazumdar, Manfrinetti, and Palenzona}}]{12}
\bibinfo{author}{\bibfnamefont{S.~K.} \bibnamefont{Dhar}},
  \bibinfo{author}{\bibfnamefont{S.}~\bibnamefont{Singh}},
  \bibinfo{author}{\bibfnamefont{P.}~\bibnamefont{Bonville}},
  \bibinfo{author}{\bibfnamefont{C.}~\bibnamefont{Mazumdar}},
  \bibinfo{author}{\bibfnamefont{P.}~\bibnamefont{Manfrinetti}},
  \bibnamefont{and}
  \bibinfo{author}{\bibfnamefont{A.}~\bibnamefont{Palenzona}},
  \bibinfo{journal}{Physica B} \textbf{\bibinfo{volume}{312-313}},
  \bibinfo{pages}{846} (\bibinfo{year}{2002}).

\bibitem[{\citenamefont{Kosaka et~al.}(1998)\citenamefont{Kosaka, Onodera,
  Ohoyama, Ohashi, , Yamaguchi, Nakamura, Goto, Kobayashi, and Ikeda}}]{13}
\bibinfo{author}{\bibfnamefont{M.}~\bibnamefont{Kosaka}},
  \bibinfo{author}{\bibfnamefont{H.}~\bibnamefont{Onodera}},
  \bibinfo{author}{\bibfnamefont{K.}~\bibnamefont{Ohoyama}},
  \bibinfo{author}{\bibfnamefont{M.}~\bibnamefont{Ohashi}}, ,
  \bibinfo{author}{\bibfnamefont{Y.}~\bibnamefont{Yamaguchi}},
  \bibinfo{author}{\bibfnamefont{S.}~\bibnamefont{Nakamura}},
  \bibinfo{author}{\bibfnamefont{T.}~\bibnamefont{Goto}},
  \bibinfo{author}{\bibfnamefont{H.}~\bibnamefont{Kobayashi}},
  \bibnamefont{and} \bibinfo{author}{\bibfnamefont{S.}~\bibnamefont{Ikeda}},
  \bibinfo{journal}{Phys. Rev. B} \textbf{\bibinfo{volume}{58}},
  \bibinfo{pages}{6339} (\bibinfo{year}{1998}).

\bibitem[{14()}]{14}
\bibinfo{note}{Note added in Proof: Very preliminary $\mu$SR results confirm
  antiferromagnetic ordering below 6.5 K, while no evidence for magnetic
  ordering was observed between 11 K and 7 K.}

\bibitem[{\citenamefont{Yamamoto et~al.}(2004)\citenamefont{Yamamoto, Takeda,
  Koyama, Mito, Wada, Shirotani, and Sekine}}]{15}
\bibinfo{author}{\bibfnamefont{A.}~\bibnamefont{Yamamoto}},
  \bibinfo{author}{\bibfnamefont{J.}~\bibnamefont{Takeda}},
  \bibinfo{author}{\bibfnamefont{T.}~\bibnamefont{Koyama}},
  \bibinfo{author}{\bibfnamefont{T.}~\bibnamefont{Mito}},
  \bibinfo{author}{\bibfnamefont{S.}~\bibnamefont{Wada}},
  \bibinfo{author}{\bibfnamefont{I.}~\bibnamefont{Shirotani}},
  \bibnamefont{and} \bibinfo{author}{\bibfnamefont{C.}~\bibnamefont{Sekine}},
  \bibinfo{journal}{Phys. Rev. B} \textbf{\bibinfo{volume}{70}},
  \bibinfo{pages}{220402(R)} (\bibinfo{year}{2004}).

\bibitem[{\citenamefont{Kosaka et~al.}(2005)\citenamefont{Kosaka, Kato, Araki,
  Môri, Nakanishi, Yoshizawa, Ohoyama, Martin, and Tozer}}]{16}
\bibinfo{author}{\bibfnamefont{M.}~\bibnamefont{Kosaka}},
  \bibinfo{author}{\bibfnamefont{Y.}~\bibnamefont{Kato}},
  \bibinfo{author}{\bibfnamefont{C.}~\bibnamefont{Araki}},
  \bibinfo{author}{\bibfnamefont{N.}~\bibnamefont{Môri}},
  \bibinfo{author}{\bibfnamefont{Y.}~\bibnamefont{Nakanishi}},
  \bibinfo{author}{\bibfnamefont{M.}~\bibnamefont{Yoshizawa}},
  \bibinfo{author}{\bibfnamefont{K.}~\bibnamefont{Ohoyama}},
  \bibinfo{author}{\bibfnamefont{C.}~\bibnamefont{Martin}}, \bibnamefont{and}
  \bibinfo{author}{\bibfnamefont{S.~W.} \bibnamefont{Tozer}},
  \bibinfo{journal}{J. Phys. Soc. Jpn.} \textbf{\bibinfo{volume}{74}},
  \bibinfo{pages}{2413} (\bibinfo{year}{2005}).

\end{thebibliography}

\end{document}